REVIEW MANUSCRIPT

# Molecular techniques employed in CTG(Ser1) and CTG(Ala) D-xylose metabolizing yeast clades for strain design and industrial applications

Ana Paula Wives[1,2,†], Isabelli Seiler de Medeiros Mendes[1,2,†], Sofia Turatti dos Santos[1,2,†], and Diego Bonatto[1,2,*]

[1]Laboratório de Biologia Molecular e Computacional, Centro de Biotecnologia da UFRGS, Departamento de Biologia Molecular e Biotecnologia, Universidade Federal do Rio Grande do Sul, Porto Alegre, RS, Brazil.

[2]Bioprocess and Biotechnology for Food Research Center (Biofood), Food Science and Technology Institute (ICTA), Universidade Federal do Rio Grande do Sul, Av. Bento Gonçalves 9500, Porto Alegre, RS, 91501-970, Brazil.

**Short title:** Molecular techniques for D-xylose fermenting yeasts.

**\*Corresponding author:**
Diego Bonatto
Centro de Biotecnologia da UFRGS - Laboratory 212
Departamento de Biologia Molecular e Biotecnologia
Universidade Federal do Rio Grande do Sul – UFRGS
Avenida Bento Gonçalves 9500 - Prédio 43421
Caixa Postal 15005
Porto Alegre – Rio Grande do Sul
BRAZIL
91509-900
Phone: (+55 51) 3308-9410
E-mail: diego.bonatto@ufrgs.br
Contract/grant sponsor: CNPq, FAPERGS
[†] These authors contributed equally to the work.

**Abstract**

D-xylose is the second most abundant monosaccharide found in lignocellulose and is of biotechnological importance for producing second-generation ethanol and other high-value chemical compounds. The conversion of d-xylose to ethanol is promoted by microbial fermentation, mainly by bacteria, yeasts, or filamentous fungi. Among yeasts, species belonging to the CTG(Ser1) or CTG(Ala) clade display a remarkable ability to ferment D-xylose to ethanol and other compounds; however, these yeasts are not employed on an industrial scale owing to their poor fermentative performance compared to conventional yeasts, such as *Saccharomyces cerevisiae,* and also because of the lack of a molecular toolbox for the development of new strains tailored to fermentation stress tolerance and performance. Thus, the purpose of this review was to evaluate the major molecular tools (e.g., transformation markers and techniques, vectors, regulatory sequences, and gene editing techniques) available for the most studied yeasts of the CTG(Ser1) clade, such as *Scheffersomyces*, *Spathaspora*, *Candida* and *Yamadazyma* species, and the CTG(Ala) clade, representative *Pachysolen tannophilus*. Furthermore, we synthesized state-of-the-art molecular developments and perspectives to design D-xylose-fermenting yeast strains.

**Introduction**

D-xylose is a major pentose found in lignocellulosic biomass and is composed of a xylan polysaccharide found in the hemicellulose fraction of lignocellulose [1,2]. It corresponds to 5–35% of the total carbohydrates found in lignocellulose biomass, depending on the plant source [1], and can be broadly classified into softwood (gymnosperms) and hardwood (angiosperms). In addition to hemicellulose, spent sulfite liquor (SSL), a byproduct of the cellulose acidic sulfite pulping process, contains 15–50% D-xylose and many other compounds, such as pentoses, hexoses, furfural and its derivatives, organic acids, and lignosulfonates [2,3]. The amount of D-xylose found in SSL depends on the wood type; softwood SSL contains a higher proportion of hexoses (e.g., glucose) than hardwood SSL, which is rich in D-xylose [3].

Considering its industrial importance, D-xylose has broad applications, ranging from biofuel production to value-added chemicals in food, health products, fuel additives, and drug manufacturing [4]. However, converting D-xylose to different chemicals by microbial fermentation is challenging, and only a few yeast species can naturally metabolize D-xylose in addition to genetically modified strains of the yeast *Saccharomyces cerevisiae* [5–7]. In some countries, softwood SSL has been used to produce second-generation bioethanol and additives for feedstock and civil construction [3]. However, untreated hardwood SSL is still not feasible for industrial applications because of the presence of high concentrations of inhibitory molecules for microbial cell growth and metabolism (e.g., acetic acid, furfural, and phenols) [5,8,9]. To solve this problem, different methods have been developed to reduce the formation of inhibitory molecules in hardwood SSL and other lignocellulose derivatives, including the addition of chemical compounds, such as inorganic hydroxides and reducing agents (e.g., sulfite), or the use of liquid-liquid and/or solid-liquid extraction procedures, thermal treatment, and biological agents (e.g., laccase and filamentous fungi) [8]. Among the methods employed, there is an increase in the fermentability of treated hardwood SSL and lignocellulose derivatives [8]. However, many of these methods require additional steps and/or treatment conditions that make their use economically unfeasible on a large industrial scale [3].

Therefore, an ideal yeast species that can tolerate inhibitory molecules present in SSL and lignocellulose derivatives with highly efficient D-xylose metabolism is desirable, and different strategies have been employed to obtain such yeast strains and species. An initial approach is to investigate yeast biodiversity in biomes, ecosystems, and ecological niches. In this approach, D-xylose-metabolizing yeast species have been isolated and characterized, such as *Pachysolen tannophilus,* which was initially present in *Castanea vesca* (sweet chestnut) and *Acacia mearnsii* (*A. mollissima;* black wattle) extracts used in leather tanning [10,11], and displayed high tolerance to the inhibitory molecules found in hardwood SSL [12]. Another characterized yeast genus isolated from the gut of the wood-boring beetle *Odontotaenius disjunctus* (Coleoptera: Passalidae) and *Phrenapates bennetti* (Coleoptera: Tenebrionidae) is *Spathaspora* sp., which can metabolize D-xylose under anaerobic conditions [13]. These yeast species are defined as "non-conventional yeasts" since they display genetic and phenotypic traits that differ from the "conventional yeasts," such as *Saccharomyces cerevisiae* [6], and constitute an unexplored source of genes and biochemical pathways with interest in industrial applications. On the other hand, the use of conventional yeasts for D-xylose fermentation has several advantages over non-conventional yeasts, especially in terms of metabolism, physiology, fermentative predictability, stress tolerance, and a large set of molecular tools available for genetic modifications and breeding selection of new strains [1,14,15]. In *Saccharomyces cerevisiae,* the enzymes of two major xylose assimilation pathways can be heterologously expressed: the oxidoreductase and isomerase pathways [14,16]. Non-conventional yeasts that naturally metabolize D-xylose employ the oxidoreductase pathway, which depends on the activity of two enzymes, xylose reductase (XR, EC 1.1.1.30) and xylitol dehydrogenase (XDH, EC 1.1.1.9), which convert D-xylose to xylulose via xylitol [14]. Bacteria (e.g., *Escherichia coli*) and chytrid fungi (e.g., *Piromyces* sp. and *Orpinomyces* sp.) mainly employ the xylose isomerase (XI, EC 5.3.1.5) pathway [14,17]. To date, the generation of recombinant *S. cerevisiae* strains able to fully convert D-xylose to ethanol by overexpression of XR-XDH or XI enzymes has only been partially achieved for a variety of reasons, including low enzyme activity, improper recombinant protein folding, metabolic imbalance (mainly due to the different requirements of XR-XDH enzymes for cofactors such as $NAD(P)H/NAD(P)^+$), accumulation of xylitol as a by-product, restrained cell growth, and lack of proper D-xylose transporters, among others [14,15]. Considering the bottlenecks associated with the heterologous expression of oxidoreductase or isomerase pathways in *S. cerevisiae,* the use of non-conventional yeast species naturally adapted to D-xylose as a carbon source appears to be a logical pathway to explore in terms of industrial production of ethanol. Therefore, the use of different molecular tools to generate new non-conventional yeast strains is essential. This review focuses on the molecular tools available for the most studied non-conventional yeast species for the CTG(Ser1) or CTG(Ala) clade employed for second-generation bioethanol production from D-xylose, such as *Scheffersomyces* stipitis, *Spathaspora passalidarum, Candida intermedia, Yamadazyma tenuis,* and *Pachysolen tannophilus*.

### *Scheffersomyces stipitis*

The yeast genus *Scheffersomyces* includes some important species known to efficiently convert D-xylose to ethanol, including *Scheffersomyces stipitis* (initially classified as *Pichia stipitis*), *Scheffersomyces shehatae, Scheffersomyces parashehatae,* and *Scheffersomyces illinoinensis* [13,18]. These yeast species are found in symbiosis within the gut of the wood-ingesting beetles, *Odontotaenius disjunctus* (Coleoptera: Passalidae) and *Verres sternbergianus* (Coleoptera: Passalidae). Characteristically, the guts of wood-ingesting beetles are oxygen-limited and have high amounts of partially degraded wood, leading to the evolution of efficient biochemical pathways to process D-xylose and complex polysaccharides in *Scheffersomyces stipitis* as indicated by complete genome sequencing of this yeast in 2007 [18].

*Scheffersomyces* belongs to the fungal CTG(Ser1) clade, which comprises other biotechnologically relevant yeast species, such as *Spathaspora passalidarum,* and is characterized by a CTG codon that encodes serine instead of leucine [19]. The reassignment of the CTG codon in the fungal CTG(Ser1) clade imposes challenges in terms of coding sequence translation, particularly for the generation of functional transformation markers commonly used for non-conventional yeast selection, such as hygromycin B (*hygR*), nourseothricin (*natR*), geneticin (*kamR*), and phleomycin (*bleR*). The coding sequence for these transformation selection markers requires codon optimization to generate a functional protein in *Scheffersomyces* sp., which was performed by Matanović et al. for *S. stipitis* [20] (Table 1). In this sense, two antibiotic-resistance markers were optimized for *S. stipitis* selection, namely *kamR* and *bleR*, with *S. stipitis* partially resistant to phleomycin and geneticin [21,22] (Table 1). In addition, the *hygR* marker was codon-optimized and proven effective in *S. stipitis* and other D-xylose-fermenting yeasts [23]. Additionally, a series of auxotrophic *S. stipitis* strains were obtained for uracil (*ura3*), leucine (*leu2*), histidine (*his3-1*), and tryptophan (*trp5-10*) biosynthesis, allowing the selection of transformed cells with vectors harboring the respective prototrophic complementation markers (Table 1) in culture media lacking specific amino acids and nitrogen bases [22,24–26]. Unfortunately, the number of auxotrophic *S. stipitis* strains available is limited compared with that of conventional yeasts. However, it is expected that introducing gene editing or DNA recombination tools could increase the number of auxotrophic markers in *S. stipitis.* In this sense, adapting a *lox*P/Cre recombination system in *S. stipitis* allows for the recycling of antibiotic resistance markers [22], which can also be applied to generate new single or combined auxotrophic markers.

Regarding the development of vectors for *S. stipitis,* an early study described a vector containing a 2 μm replicon from *S. cerevisiae* and a non-optimized *kamR* sequence (pUCKm8 vector; Table 2) that could replicate in *S. stipitis* [21]. However, the authors did not evaluate the vector stability. Matanović et al. and Yang et al. [20,26] described plasmids containing antibiotic resistance or prototrophic (*URA3*) markers with an autonomously replicating sequence (*ARS2*) isolated from *S. stipitis* (pRS54 and pJM2/6 vector series; Table 2). The authors showed that these vectors could be episomally maintained for at least 50 generations or stably integrated into the genome of *S. stipitis* when linearized [26] (Table 2). Another important achievement was the development of a series of stable centromeric vectors containing a centromeric sequence derived from chromosome five (*CEN5*) in addition to an *ARS* (*ARS2*) [27]. *CEN5* increases the expression of heterologous sequences, including Cas9, which allows for better gene knockout efficiency [27,28]. Interestingly, *CEN5,* as well as other centromeric sequences from *S. stipitis,* appear to be part of the long terminal repeat (LTR) of Tps5 (a Ty5-type retrotransposon) that forms a cluster in the centromeres of *S. stipitis* [29]. Finally, integrative vectors using the *URA3* (YIPs1; Table 2) or rDNA 18S sequence from *Spathaspora passalidarum,* which contain different codon-adapted antibiotic-resistance markers (pRTH series; Table 2), were successfully employed for *S. stipitis* transformation with high stability, albeit with low efficiency [23,26].

Developing vectors for *S. stipitis* is essential for establishing a gene editing system using CRISPR-Cas9 [27,28]. However, as observed in almost all non-conventional yeasts [30], *S. stipitis* preferentially repairs DNA double-strand breaks induced by Cas9 by a non-homologous end-joining (NHEJ) DNA repair mechanism instead of homologous recombination (HR), generating mainly insertion-deletions (indels) in the DNA target molecule and hampering the use of gene editing techniques such as allele replacement [31]. A common way to circumvent the NHEJ pathway is to delete the coding sequences of the Ku complex, which is composed of Ku70 and Ku80 proteins responsible for directing the molecular components of the NHEJ pathway to the DSB lesion site [32]. The *ku70Δ/ku80Δ* technique was employed for *S. stipitis* gene editing with a high HR rate and gene knockdown expression using dead Cas9 (dCas9) coupled with a repressor factor [31]. However, a significant drawback of the *ku70Δ/ku80Δ* technique is diminished cell growth and telomere instability [30], which impair the industrial usability of *ku70Δ/ku80Δ* strains for most non-conventional yeast species. Finally, the development of the so-called "bio-luminescent indicator nullified by

Cas9-actuated recombination" (BLINCAR) allows the delivery and/or deletion of multiple genetic elements within *S. stipitis* genome [33] using replicative and/or integrative vectors combining the Ss*ARS2* replicon or long segments (≥ 2.0 kb) of amplified target genome sequences with different antibiotic resistance markers, like *hygR* (pWR vector series; Table 2).

Efficient transformation techniques are fundamental for the development of new strains of *Scheffersomyces* sp. and other D-xylose-fermenting yeasts. An early successful electroporation protocol for *S. stipitis* was designed by Ho et al. [21] using the pUCKm8 vector (Tables 2 and 3); however, transformation efficiency was not evaluated. Further, an electroporation efficiency of $6.0 \times 10^2$ to $8.6 \times 10^3$ colony-forming units (CFU) per ug of DNA was achieved using the pJM2/6 vector series (Tables 2 and 3) [26]. Cao et al. [31] improved electroporation efficiency by increasing the voltage to 2.5 kV, obtaining $1.0 \times 10^3$ CFU for 1 μg of a 10 kb plasmid. In addition to electroporation, *Scheffersomyces* employed a modified lithium acetate/polyethylene glycol (LiAc/PEG) transformation procedure, although the transformation efficiency was not indicated (Table 3) [33,34].

Considering the development of DNA expression devices (e.g., genes), the availability of promoters and terminators for *Scheffersomyces* sp. and other non-conventional yeast species is crucial for yeast strain design. Different numbers of constitutive and inducible promoters have been described in *S. stipitis* (Table 4). Strong constitutive promoters, such as *TEF1*p and *TDH3*p, and terminators from *XYL1* and *XYL2* were isolated from *S. stipitis* and employed to express the fumarate biosynthesis pathway in *Rhizopus oryzae* [35]. In addition, *TEF1* promoters from different species, such as *Blastobotrys adeninivorans* have been shown to function in *S. stipitis* and other xylose-metabolizing yeasts, such as *Spathaspora passalidarum, Candida jeffriesii,* and *Candida amazonensis* [23]. In the same study, the authors showed that native promoter sequences from *S. stipitis XYL1.1*, *S. passalidarum ADH1,* and *S. passalidarum XYL1.1* could drive the expression of *EGFP* in their respective hosts, but not necessarily in different D-xylose-fermenting yeasts, pointing to genera- or species-specific transcriptional regulatory mechanisms [23]. Similarly, terminator sequences from D-xylose-fermenting yeasts are not necessarily interchangeable among different species; however, *CYC1*t from *S. passalidarum* promotes gene expression in *S. stipitis* [23]. Additionally, the derived promoter and terminator sequences from *TEF1* of *Eremothecium gossypii* have been proven to work on *S. stipitis* strains and other xylose-fermenting yeasts [20]. Finally, the inducible promoters *GAL1* and *XKS1* (Table 4) from *S. stipitis* could be alternatives for regulating the expression of specific genes in the presence of galactose and D-xylose, respectively [33].

### *Spathaspora passalidarum*

The genus *Spathaspora* contains 22 described yeast species that use D-xylose from lignocellulosic residues to produce ethanol and/or xylitol with high efficiency under aerobic or anaerobic conditions, even in the presence of glucose and cellobiose [13,36]. The most characterized species of *Spathaspora* is *S. passalidarum,* isolated from the wood-boring beetle *O. disjunctus* (Coleoptera: Passalidae) [37], followed by *S. arborariae,* which was isolated from rotting wood in the Atlantic Rain Forest and Cerrado ecosystem in Brazil [38]. As they belong to the CTG(Ser1) clade [19], the molecular tools available for *Spathaspora* sp. are even more limited than those for *S. stipitis;* however, they are more or less interchangeable between both yeast genera, especially antibiotic-resistant transformation markers (Table 1) [23]. To date, no studies have described the use of prototrophic and/or dominant markers for auxotrophic *S. passalidarum* strains (Table 1).

Different vectors optimized for *Spathaspora* are also scarce. An integrative vector containing the rDNA 18S sequence from *S. passalidarum* and codon-adapted *hygR* was used to express *EGFP* (pRTH vector series; Table 2) with a low transformation efficiency [23]. In addition, an epissomal vector containing a non-optimized codon *kamR* marker and a 2 µm replicon from *S. cerevisiae* (YEp352-km vector; Table 2) was shown to replicate in *S. passalidarum* [39].

Two lithium acetate transformation protocols were developed for *S. passalidarum* (Table 3). One protocol employed the LiAc/PEG/ssDNA method using one microgram of the linearized integrative vector series pRTH (Table 2), followed by incubation at 30 °C for 3 h in a rich (YPD) medium [23], and the second employed a LiAc/PEG/single-stranded DNA (ssDNA) optimized method for the YEp352-km vector (Table 2) [39]. Both protocols can be employed for *S. passalidarum* transformation, albeit with low efficiency (31–50 transformants/µg of pRTH vector for the first protocol and approximately 170 transformants/µg of YEp352-km vector under optimized conditions for the second protocol; Table 3) [39]. In addition, to date, no electroporation or trans-kingdom conjugation protocols have been described for *Spathaspora* species and/or strains (Table 3).

To our knowledge, no gene editing tools have been developed or adapted for *Spathaspora* species. Finally, promoters from *S. passalidarum ADH1, XYL1.1,* and *E. gossypii TEF1* can drive heterologous *EGFP* expression in this yeast (Table 4) [20,23]. Moreover, different terminator sequences, mainly derived from *S. passalidarum* genes (and the *S. cerevisiae CYC1* terminator), were proven to work on this yeast (Table 4).

Interestingly, the *S. passalidarum XYL1.1* terminator does not work on *S. stipitis* [23], indicating that not all DNA components are interchangeable between *S. passalidarum* and *S. stipitis.*

### *Candida intermedia* and *Yamadazyma tenuis*

The genus *Candida* represents a large group that contains important yeast species, such as pathogenic *C. albicans, C. glabrata, C. parapsilosis, C. tropicalis,* and *C. krusei* [40]. Many *Candida* species can metabolize D-xylose, including *C. insectorum* and *C. jeffriesii* (both isolated from wood-ingesting insects), *C. ligosa* and *C. tropicalis* (isolated from rotting wood), and *C. famata* and *C. guilliermondii* (isolated from fruits) [41]. In addition to *Candida,* the *Yamadazyma* monophyletic clade was described as a sister group of *Candida* isolated from tree bark and is capable of metabolizing D-xylose [42,43]. Two species of *Candida* and *Yamadazyma* have been well studied for D-xylose fermentation to bioethanol, namely *C. intermedia* and *Y. tenuis* (formerly known as *Candida tenuis*) [44]. Similar to *Scheffesomyces* and *Spathaspora*, *C. intermedia* and *Y. tenuis* belong to the CTG(Ser1) clade [43,45]. Moreover, molecular tools widely available for *Candida* (mainly designed for *C. albicans*) can potentially be employed for *Yamadazyma* owing to their phylogenetic proximity to *Candida* clade.

*C. intermedia* has also been used for xylitol and single-cell proteins from D-xylose [46], has an efficient D-xylose transport system [47], and can metabolize glucose and D-xylose at high concentrations [48]. A few molecular tools have been developed for *C. intermedia,* including a split-marker deletion approach for gene editing based on vectors containing the *C. albicans* dominant marker nourseothricin acetyltransferase (*CaNAT1*) (Tables 1 and 2) flanked by 1 kb sequences with homology to the target sequence (template DNA) [45]. The authors employed electroporation for vector transformation with low efficiency (Table 3). Using this approach, the authors improved the HR rates of target sequence editing to 70% [45]. However, we could not find any other reports indicating the use of precision gene-editing techniques for *C. intermedia* and *Y. tenuis,* such as Cas9. Despite this, the use of Cas9 technology for *C. albicans* and other *Candida* species is well established and can be adapted for *C. intermedia* and *Y. tenuis* [49]. Similarly, promoters, terminators, transformation markers, different transformation techniques, and replicative plasmids containing *ARS* or integrative plasmids that are available for *C. albicans, C. tropicalis, C. glabatra, and C. maltosa* could potentially be used for *C. intermedia* and *Y. tenuis* [50,51]. The bovine serum albumin (BSA)-responsive promoter (*SAP2*) of *C. albicans* has been shown to regulate the expression of a recombinant flippase (FLP) in *C. intermedia* in culture media containing BSA as a unique nitrogen source [45], reinforcing the idea that the molecular tools available for *C. albicans* can be directly employed for *C. intermedia* genetic manipulation. Finally, we did not find any molecular tools for *Y. tenuis* (Tables 1 to 4).

***Pachysolen tannophilus***

*Pachysolen tannophilus* is a D-xylose-fermenting model yeast belonging to the CTG(Ala) clade [52]. However, molecular tools for the genetic analysis and/or modification of *P. tannophilus* have been less developed than other D-xylose-fermenting yeasts, such as *S. stipitis.* Considering transformation markers, *P. tannophilus* auxotrophs for different amino acid requirements have been generated by UV mutagenesis [53–55], where *leu*[-] strains are conventionally employed for transformation with plasmids containing the *LEU2* gene from *S. cerevisiae* (Table 1) because its coding sequence contains only the CTG codon and does not interfere with the activity of the

Leu2 protein [56]. However, similar to yeasts from the CTG(Ser1) clade, antibiotic-resistance transformation marker coding sequences should have their codons optimized for *P. tannophilus* (e.g., changing the CTG codon to TTG codon) because the codon CTG codifies to Ala instead of Leu in this yeast [56]. An early attempt to express the *kamR* transformation marker in *P. tannophilus* was unsuccessful, probably because of the non-optimized *kamR* coding sequence [57]. Furthermore, it was shown that even codon-optimized *kamR* and aureobasidin A resistance genes (*AUR1-C*) were unable to confer resistance to both geneticin and aureobasidin A in *P. tannophilus* under the control of native *S. cerevisiae LEU2* and *AUR1* promoters (Table 1) [56], pointing that *S. cerevisiae* promoters negatively affect the expression of these markers. Reinforcing this idea, a codon-optimized hygromycin B-resistance (hygR) coding sequence controlled by *P. tannophilus* native glyceraldehyde-3-phosphate dehydrogenase promoter and terminator has been shown to work in *P. tannophilus* [58]. Interestingly, *S. cerevisiae* hexokinase isoenzyme 2 (*HXK2*) was employed as a dominant transformation marker in hexose-metabolizing negative mutants of *P. tannophilus* (Table 1) [59].

Very few vectors are available for *P. tannophilus* (Table 2). Epissomal vectors containing the *S. cerevisiae* 2 μm replicon combined with the prototrophic *S. cerevisiae LEU2* marker (YEp13 or pACT2-Cas9c vectors; Table 2) or replicative vectors containing the *ARS* from *S. cerevisiae* combined with the dominant marker *HXK2* (YRp vector; Table 2) have been used for *P. tannophilus* transformation [51,56], but no data have been reported on plasmid stability in this yeast, and no integrative vectors have been reported to date for *P. tannophilus*. Additionally, the use of a centromeric sequence from *S. cerevisiae* (*CEN4*) and *ARS1* from the same yeast proved ineffective for plasmid replication and stability in *P. tannophilus* (Table 2), indicating that the design of centromeric vectors depends on the characterization of centromeres and *ARS* in *P. tannophilus* [56].

Different transformation methods based on the use of lithium acetate (LiAc) alone [59] or combined with ssDNA and PEG have been employed for *P. tannophilus* (Table 3) [56]. The efficiency of LiAc transformation was reported to be 40-50 transformant/μg vector [59], whereas the combined LiAc-ssDNA-PEG technique achieved an efficiency of $3 \times 10^4$ transformants/μg DNA (Table 3) [56]. Interestingly, trans-kingdom conjugation between *P. tannophilus* and *Escherichia coli* has been applied with relative success to promote the transfer of vectors [53,56], with $2.99 \times 10^{-2}$ transconjugants per recipient at a donor-to-recipient ratio of 1000:1 (Table 3) [56]. To date, no electroporation technique has been reported in *P. tannophilus*.

Although this yeast has been known since 1957 [11], little has been done to prospect and/or characterize regulatory sequences (e.g., promoters and terminators) for heterologous gene expression or gene reengineering, except for the promoter and terminator of *P. tannophilus GAPDH* gene (Table 4), which has been proven to be efficient in driving the expression of *hygR* [58]. Moreover, little effort has been made to develop transformation markers, vectors, and gene-editing techniques for this yeast species. Considering that *P. tannophilus* belongs to a few groups of yeasts capable of tolerating the toxic components of hardwood SSL [12] and can convert glycerol to ethanol [60], the development of molecular tools could improve the biotechnological importance of this yeast for biofuel production.

**Conclusion**

The D-xylose-fermenting yeasts of the CTG(Ser1) and CTG(Ala) clades are an important source of genotypes and phenotypes useful for converting lignocellulosic biomass to biofuels and other industrially important chemical compounds. However, many of these yeast species are still unsuitable for industrial applications because of the lack of knowledge about their physiology and molecular tools that enable researchers to develop new strains tailored for industrial applications compared to conventional yeasts (e.g., *S. cerevisiae*). A major barrier to the development of molecular tools for these yeasts is the reassignment of the CTG codon for Ser or Ala, which impairs the use of existing vectors, transformation markers, regulatory sequences, and precise gene editing technologies available for conventional yeasts. Moreover, highly efficient transformation techniques and characterization of centromeres and autonomously replicative sequences for yeasts of the CTG(Ser1) and CTG(Ala) clades for the development of stable vectors still lag behind those observed for conventional yeasts. Considering that synthetic biology technologies, such as gene synthesis, cloning, and genome annotation tools, are advancing rapidly, the development of new molecular tools is expected to increase, allowing the design of new strains adapted for industrial applications.

**Funding**

This work was supported by "CNPq/MCTI/CT-BIOTEC N° 30/2022 - Linha 2: Novas tecnologias em Biotecnologia" from Conselho Nacional de Desenvolvimento Científico e Tecnológico – CNPq [grant number 440226/2022-8], by "Programa Pesquisador Gaúcho - PqG" from Fundação de Amparo à Pesquisa do Estado do Rio Grande do Sul - FAPERGS [Edital FAPERGS 07/2021 and grant number 21/2551-0001958-1], by "INOVA CLUSTERS TECNOLÓGICOS" from Fundação de Amparo à Pesquisa do Estado do Rio Grande do Sul - FAPERGS [Edital FAPERGS 02/2022 and grant number 22/2551-0000834-8] and by "Programa de Redes Inovadoras de Tecnologias Estratégicas do Rio Grande Do Sul – RITEs-RS" from Fundação de Amparo à Pesquisa do Estado do Rio Grande do Sul - FAPERGS [Edital FAPERGS 06/2021 and grant number 22/2551-0000397-4]. The sponsors had no role in the study design, collection, analysis, interpretation of data, writing of the report, or decision to submit the article for publication.

**Authors contributions**

**Ana Paula Wives:** Writing - Review & Editing, Visualization. **Isabelli Seiler de Medeiros Mendes:** Writing - Review & Editing, Visualization. **Sofia Turatti dos Santos:** Writing - Review & Editing, Visualization. **Diego Bonatto:** Conceptualization, Writing - Original Draft, Writing - Review & Editing, Visualization, Supervision, Project administration, Funding acquisition.

**Declaration of interest**

Declarations of interest: none.

**Tables**

Table 1. Transformation markers employed in CTG(Ser1) and CTG(Ala) clade yeasts.

| Yeast species | Transformation markers | | |
|---|---|---|---|
| | **Prototrophic markers** | **Antibiotic-resistance markers** | **Other marker (dominant markers)** |
| *Scheffersomyces stipitis* | *URA3; LEU2; HIS3-1; TRP5-10* [22,24–26] | *kamR*[1,2]; *bleR*[1,2]; *natR*[1]; *hygR*[1] [20–23] | Not reported |
| *Spathaspora passalidarum* | Not reported | *kamR*[1,2]; *bleR*[1,2]; *natR*[1]; *hygR*[1] [23] | Not reported |
| *Candida intermedia* | Not reported | *Candida albicans natR* [45] | Not reported |
| *Yamadazyma tenuis* | Not reported | Not reported | Not reported |
| *Pachysolen tannophilus* | *SceLEU2* [56] | 1. Non-optimized codon sequence of kamR do not confer resistance to geneticin [57]<br>2. Codon-optimized *kamR* and *SceAUR1-C* were observed not to confer resistance to geneticin and aureobasidin A, respectively [56].<br>3. Codon-optimized *hygR* confers resistance to hygromycin B [58] | *SceHXK2* [59] |

[1]Requires codon optimization.

[2]*Scheffersomyces stipitis* is partially resistant to phleomycin and geneticin (up to 2,000 µg/mL for geneticin) [23,33]

Abbreviation: *Saccharomyces cerevisiae* (*Sce*)

Table 2. Different vector types employed for the transformation of D-xylose fermenting yeasts. The table indicates the replicon/integration type (boldface) and transformation markers. In addition, the names of vectors are shown inside parentheses.

| Yeast species | Vector type | | | |
|---|---|---|---|---|
| | **Epissomal** | **Replicative** | **Centromeric** | **Integrative** |
| Scheffersomyces stipitis | 1. **S. cerevisiae 2 µ replicon**; kamR (pUCKm8 vector) [21] | 1. **SsARS2**; antibiotic resistance markers[1] (pRS54 series and pWR series) [20,33]<br>2. **SsARS2**; SsURA3 (pJM2/6 series) [26] | 1. **SsCEN5; SsARS2**; SsURA3 [28] | 1. **SsURA3**[2] (YIPs1) [26]<br>2. **Spa18S**[3] rDNA; hygR (pRTH series) [23]<br>3. **SsURA3**[4], **SsEGC3**[4], **SsGSC2**[4], **and SsCRF1**[4]; antibiotic resistance markers[1] (pWR series) [33] |
| Spathaspora passalidarum | 1. **S. cerevisiae 2 µ replicon**; kamR (YEp352-km vector) [39] | Not reported | Not reported | 1. **Spa18S**[3] rDNA; hygR (pRTH series) [23] |
| Candida intermedia | Not reported | Not reported | Not reported | 1. **CiADE2**[5]; **CiXYL1_2C**[5]; **CiLAC9**[5]; **CiLADA**[5]; C. albicans natR[5] [45] |
| Yamadazyma tenuis | Not reported | Not reported | Not reported | Not reported |
| Pachysolen tannophilus | 1. **S. cerevisiae 2 µ replicon**; SceLEU2 (YEp13) [53]<br>2. **S. cerevisiae 2 µ replicon**; | 1. **S. cerevisiae ARS**; SceHXK2 (YRp) [59] | 1. **S. cerevisiae CEN4-ARS1** unable to replicate in P. tannophilus (pAUR123 and derivatives) | Not reported |

| Yeast species | Vector type | | | |
|---|---|---|---|---|
| | *SceLEU2* (pACT2-CAS9c) [56] | | [56] | |

[1]Transformation markers: *kamR*; *bleR*; *natR*; *hygR*.

[2]The ~ 4.1 kb *SsURA3* sequence was cut with different restriction endonucleases and employed for *S. stipitis* TJ26 (*ura3-1*) electroporation. Southern blot data showed that *SsURA3* genome integration was random due to non-homologous end-joining recombination (NHEJ).

[3]The ~1.55 kb *Spa*18S rDNA sequence was cut with *Stu*I, and the linearized vector was employed for yeast transformation.

[4]A 2.7 kb genomic segment containing the *SsURA3* gene was employed for vector integration (pWR58 vector) after cutting with *Eco*RV. The 2 kb genomic sequences containing *SsEGC34, SsGSC24,* and *SsCRF14* were amplified by PCR, cloned into the pWR88 vector, and linearized with *Eco*RV.

[5]Two vectors were generated in this study. Vector p*CaNAT1* and its derivative with flippase (FLP) recombinase (p*FLP_CaNAT1*). Both vectors contained the *Candida albicans natR* transformation marker under the control of the *EgTEF1* promoter and terminator, while *CaSAP2* promoter induces FLP recombinase. A homology sequence of ~1.0 kb on both sides of the indicated target genes was used for genome integration.

Abbreviations: *Blastobotrys adeninivorans* (*Ba*); *Eremothecium gossypii* (*Eg*); *Pachysolen tannophilus* (*Pt*); *Saccharomyces cerevisiae* (*Sc*); *Scheffersomyces stipitis* (*Ss*), *Spathaspora passalidarum* (*Sp*).

Table 3. Transformation techniques described for D-xylose fermenting yeasts. When available, the transformation efficiency is indicated inside parentheses.

| Yeast species | Transformation technique | | |
|---|---|---|---|
| | **Chemical** | **Electroporation** | **Trans-kingdom conjugation** |
| *Scheffersomyces stipitis* | Yes (LiAc/PEG/ssDNA) - no reported efficiency [33,34] | Yes ($6.0 \times 10^2$ to $8.6 \times 10^3$ CFU per ug of DNA) [21,26,31] | Not reported |
| *Spathaspora passalidarum* | Yes<br>1. LiAc/PEG/ssDNA: 3.1 to $5.0 \times 10^1$ CFU per ug of DNA [23]<br>2. LiAc/PEG/ssDNA: ~ $1.7 \times 10^2$ CFU per ug of DNA [39] | Not reported | Not reported |
| *Candida intermedia* | Not reported | Yes ($7.8 \times 10^1$ CFU per ug of DNA) [45] | Not reported |
| *Yamadazyma tenuis* | Not reported | Not reported | Not reported |
| *Pachysolen tannophilus* | Yes (LiAc): $4.0 - 5.0 \times 10^1$ CFU per ug of DNA [59]<br>(LiAc/PEG/ssDNA): $3.0 \times 10^4$ CFU per ug of DNA [56] | Not reported | Yes ($2.99 \times 10^{-2}$ transconjugants per recipient at donor-to-recipient ratio of 1000:1) [53,56] |

Abbreviations: colony-forming units (CFU); lithium acetate (LiAc); polyethylene glycol (PEG); single-stranded DNA (ssDNA).

Table 4. Regulatory sequences (promoters and terminators) used for the expression of coding sequences in D-xylose fermenting yeasts. Inducible promoter sequences are shown in boldface.

| Yeast species | Regulatory sequences | |
|---|---|---|
| | **Promoters** | **Terminators** |
| *Scheffersomyces stipitis* | *SsTEF1; SsTDH3; BaTEF1; SsXYL1.1; EgTEF1; SsCCW12; SsPGK1; SsHHF1; SsADH1; SsCYC1;* **SsGAL1**[1]**;** **SsXKS1**[1] [20,23,33,35] | *SsXYL1; SsXYL2; SpCYC1;SsADH1; SsADH2; SsACT1* [20,23,33,35] |
| *Spathaspora passalidarum* | *EgTEF1; SpTEF1; SpADH1; SsXYL1.1* [23] | *ScCYC1; SpXYL1.1; SpTEF1; SpADH1* [23] |
| *Candida intermedia* | *EgTEF1;* **CaSAP2**[2] [45] | *EgTEF1* [45] |
| *Yamadazyma tenuis* | Not reported | Not reported |
| *Pachysolen tannophilus* | *PtGAPDH* [58] | *PtGAPDH* [58] |

[1]The *SsGAL1* and *SsXKS1* promoters are induced by the presence of galactose or xylose in culture media.

[2]The *CaSAP2* promoter is induced by bovine serum albumin (BSA) in culture media without additional nitrogen sources.

Abbreviations: *Blastobotrys adeninivorans* (*Ba*); *Candida albicans* (*Ca*); *Eremothecium gossypii* (*Eg*); *Pachysolen tannophilus* (*Pt*); *Saccharomyces cerevisiae* (*Sc*); *Scheffersomyces stipitis* (*Ss*), *Spathaspora passalidarum* (*Sp*).